# Anisotropic Time-Domain Electronic Response in Cuprates

F. Giusti[1,2], A. Montanaro[1,2], A. Marciniak[1,2], F. Randi[3], F. Boschini[4,5,6], F. Glerean[1,2], G. Jarc[1,2], H. Eisaki[7], M. Greven[8], A. Damascelli[4,5], A. Avella[9,10], D. Fausti[1,2] *

[1] *Department of Physics, Università degli Studi di Trieste, 34127 Trieste, Italy*
[2] *Elettra Sincrotrone Trieste S.C.p.A., 34127 Basovizza Trieste, Italy*
[3] *Joseph Henry Laboratories of Physics, Princeton University, Princeton, New Jersey 08544, USA*
[4] *Quantum Matter Institute, University of British Columbia, Vancouver, BC V6T 1Z4, Canada*
[5] *Department of Physics & Astronomy, University of British Columbia, Vancouver, BC V6T 1Z1, Canada*
[6] *Centre Énergie Matériaux Télécommunications, Institut National de la Recherche Scientifique, Varennes, Québec J3X 1S2, Canada*
[7] *Nanoelectronics Research Institute, National Institute of Advanced Industrial Science and Technology, Tsukuba, Ibaraki 305-8568, Japan*
[8] *School of Physics and Astronomy, University of Minnesota, Minneapolis, Minnesota 55455, USA*
[9] *Dipartimento di Fisica "E.R. Caianiello," Universita` degli Studi di Salerno, I-84084 Fisciano (SA), Italy*
[10] *CNR-SPIN, UOS di Salerno, I-84084 Fisciano (SA), Italy*

\* <daniele.fausti@elettra.eu>

Superconductivity in the cuprates is characterized by spatial inhomogeneity and an anisotropic electronic gap of d-wave symmetry. The aim of this work is to understand how this anisotropy affects the non-equilibrium electronic response of high-Tc superconductors. We compare the nodal and antinodal non-equilibrium response to photo-excitations with photon energy comparable to the superconducting gap and polarization along the Cu-Cu axis of the sample. The data are supported by an effective d-wave BCS model indicating that the observed enhancement of the superconducting transient signal mostly involves an increase of pair coherence in the antinodal region, which is not induced at the node.

Electronic inhomogeneity [1-10] and anisotropic superconducting (SC) gap in reciprocal space, which features a d-wave symmetry [11-14] are among the most prominent characteristics of the cuprate superconductors. In d-wave superconductors, the electronic transitions at the antinode are limited to frequencies larger than the SC gap, whereas low energy excitations are generally allowed at all frequencies at the node, where the SC gap vanishes. This feature dominates the electronic response and differentiates cuprates from conventional s-wave superconductors [15-19]. We recently reported a marked dependence of the response of cuprate superconductors on the polarization of ultrashort light pulses with photon energies comparable to the antinodal SC gap [20]. We revealed that mid-Infrared pulses at frequencies comparable to the gap energy and polarized along the Cu-Cu direction drive an enhancement of the SC pairs phase coherence.

Here we study the dynamical response of the electronic excitations and of the SC order parameter arising from different regions of the Fermi surface [21]. By selecting the polarizations of the incoming and the outgoing probe beams, we use time-domain electronic Raman scattering to probe excitations with different symmetries and, thereby, demonstrate that the SC gap exhibits a dynamic response that is different in different parts of the first Brillouin zone. In particular, we study the electronic response of optimally doped

Bi$_2$Sr$_2$Y$_{0.08}$Ca$_{0.92}$Cu$_2$O$_{8+\delta}$ (Y-Bi2212) across the SC transition following a mid-infrared photo-excitation polarized along the Cu-Cu axis and reveal that, whereas antinodal excitations depend strongly on the pump frequency, the nodal ones do not. By rationalizing this experimental evidence through an effective d-wave symmetry BCS model, our findings indicate that low frequency ac-currents along the Cu-Cu direction, driven by long wavelength mid-infrared excitations, may induce a dynamical enhancement of antinodal coherence, possibly also above T$_c$.

At optimal doping, Y-Bi2212 exhibits a bulk superconductivity below T$_c$ = 97 K, a pseudogap phase between T$_c$ and T$^*$ = 135 K, a "strange-metallic" phase above T$^*$ [22]. The SC state is characterized by a maximum SC gap amplitude of 2Δ (T=0 K)~75 meV [23]. The Bi-based cuprates are prototypical high-Tc superconductors and their phase diagram has been extensively explored via a variety of equilibrium and time-resolved techniques [22-38]. The sample studied here was grown and characterized as described in ref [39]. Y-Bi2212 structure can be approximated by a D$_{4h}$ tetragonal point group and, for parity reasons, the only Raman active modes have $A_{1g}$, $A_{2g}$, $B_{1g}$, $B_{2g}$ and $E_g$ symmetry [40].

In the experimental geometry employed here, where the beam propagation direction is perpendicular to the Cu-O planes, we define the angles $\alpha$ and $\theta$ as the polarization directions with respect to the crystallographic Cu-O axis of the input and output probe beams, respectively. In this framework we can express the matrix element of the Raman tensor as

$$R(\alpha,\theta) = \frac{1}{2}\left[R_{A1g}\cos(\alpha-\theta) + R_{B1g}\cos(\alpha+\theta) + R_{B2g}\sin(\alpha+\theta)\right] \quad (1)$$

The *B* modes can be selected through crossed polarization configurations ($\theta=\alpha \pm 90°$), whereas the total symmetric mode $A_{1g}$ cannot be singled out by linearly polarized beams, although its relative high intensity renders all other contributions negligible in parallel polarization measurements ($\alpha=\theta$) [40].

Importantly, by changing Raman scattering configurations, one gains sensitivity to different areas of the reciprocal space: while the $B_{1g}$ configuration selects an area close to the d-wave antinode (Figure 1 c), the B$_{2g}$ symmetry probes the nodal region (Figure 1e) [20].

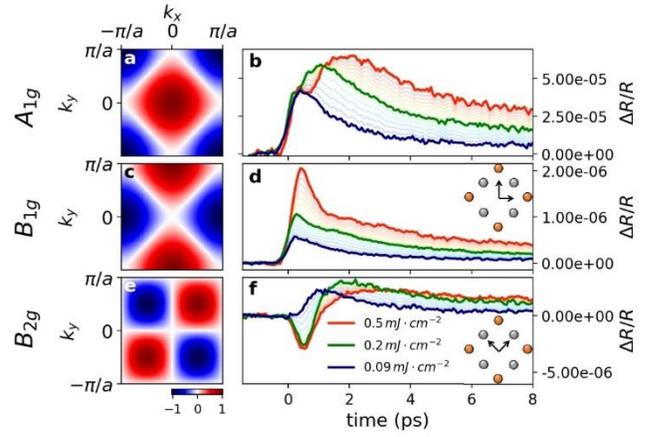

**Figure 1: Polarization selection.** Left column: sensitivity to the excitations in the first Brillouin zone for different Raman symmetries (**a** $A_{1g}$, **c** $B_{2g}$ and **e** $B_{1g}$). Right column: results of the pump-probe experiment at T=80 K (below T$_c$) for different excitation fluence. **b** No polarization selection (the dominant contribution is $A_{1g}$), **d** $B_{1g}$ and **f** $B_{2g}$. Three excitation fluences are considered: 0.09 mJ cm$^{-2}$ (blue line) corresponds to the commonly dubbed "linear" response regime, whereas 0.2 mJ cm$^{-2}$ and 0.5 mJ cm$^{-2}$ (green and red line, respectively) refer to the "non-linear" response regime associated to strong perturbation of the SC gap. The inset of panels **d** and **f** represent sketches of the Cu-O layer of cuprates (the orange dots represent the Cu atoms and the gray ones the oxygen atoms) and the polarization of the incoming and selected probe beams (arrows).

Figure 1 displays the differential reflectivity probed via ultrashort light pulses centered at 1.63 eV with mid-infrared excitation polarized along the Cu-Cu axis and different incoming and outgoing polarization selections of the probe: (**b**) polarization integrated measurements, (**d**) $B_{1g}$ and (**f**) $B_{2g}$ configurations.

These measurements were performed in the SC phase (T=80 K) in different excitation intensity regimes. The low intensity ones (blue lines) are in the "linear" response regime in which the SC order parameter is weakly perturbed. Conversely, higher pump intensities (green and red lines) access the "non-linear" regime where the strong perturbation of the SC phase causes a non-trivial composite response. In the latter case, the observed dip in Fig.1**b** at ≈0.5 ps is usually ascribed to a dynamical melting of the SC phase due to the strength of the photo-excitation [41-45]. Importantly, the time domain response for the different polarization configurations, and consequently for different momentum regions, reveals distinct non-linear dynamical processes: whereas the nodal signal (Fig. 1**f**) decreases as the pump fluence increases, together with $A_{1g}$, the response around the antinode (**d**) survives. Therefore, when the photo-excitation is intense enough to perturb the SC state, beyond the linear regime, and the negative component appears, the antinodal signal is enhanced (Fig. 1**d**); Figure 1**b** is therefore reinterpreted ad the result of the composite nodal and antinodal dynamics.

The dynamic response changes at 97 K, when the sample exits the bulk SC phase. Although the temperature dependence of the pump-probe signal has already been investigated in the cuprates [20,40], a study of the mid-infrared electronic dynamics induced for different areas of the Fermi surface is missing. In Figure 2, we show the results of the measurements for $B_{1g}$ and $B_{2g}$ configurations as a function of temperature. Excitations with energy lower than $2|\Delta|$ (where $\Delta$ is the gap amplitude at low temperature) and polarization parallel to the Cu-Cu crystallographic direction result in a peculiar gap response: Figure 2 shows the response of the system to excitation with photon energies 170 meV ($>2\Delta$) and 70 meV ($<2\Delta$) and fluence f=0.09 mJ cm$^{-2}$. The variation of the optical properties of the sample is probed by a near-infrared (hν=1.63 eV) ultrashort laser pulse.

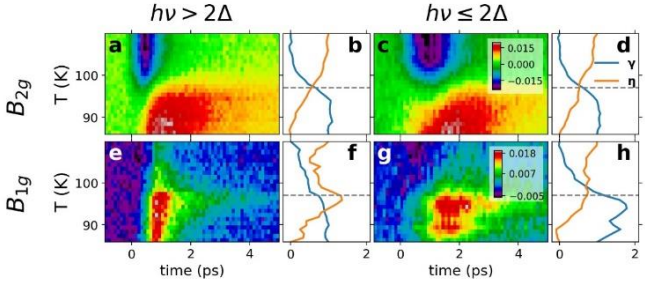

**Figure 2**: **Temperature dependence for different pump photon energies.** Temperature-dependent pump-probe measurements, in $B_{2g}$ and $B_{1g}$ configurations, at low fluence (0.09 mJ cm$^{-2}$). Top row: $B_{2g}$ measurements with photo-excitation energies of **a** 170 meV and **c** 70 meV. Panels **b** and **d** display the values of $\gamma(T)$ and $\eta(T)$ (Eq. 2) obtained from the data in **a** and **c**, respectively. The dashed grey lines highlight the critical temperature $T_c$. Bottom row (Figs. **e-h**): same analysis for measurements in the $B_{1g}$ configuration.

By visual inspection of Figure 2, we note that the "antinodal" ($B_{1g}$) temperature maps (panels **e** and **g**) strongly depend on the pump photon energy, whereas the "nodal" ($B_{2g}$) ones show no significant difference (panels **a** and **c**). In order to qualitatively quantify this difference, we fit the signal at a fixed temperature to a linear combination of a SC and a pseudogap time-resolved signal, as described by the relation

$$f(T) = \gamma(T)S_{SC} + \eta(T)S_{PG} \quad (2)$$

where $f(T)$ is the pump-probe signal at temperature $T$, and $S_{SC}=f(T=85\ K)$ and $S_{PG}=f(T=110\ K)$ are the signal measured below and above the critical temperature, respectively. In this way, the coefficients $\gamma(T)$ and $\eta(T)$ effectively weight the contributions of the superconducting and pseudogap signals to the response at a given temperature T. We stress that the qualitative results discussed here do not depend on the arbitrarily chosen temperature (as long as they are well within the two phases) and that this procedure does not aim at identifying Tc, but it is rather used to obtain a qualitative criterion to compare the response at different wavelength for the two polarization configurations (see supplemental materials for details). The results of this procedure are displayed in Fig. 2 **b, d, f, h**.

In spite of the qualitative nature of this analysis, the temperature dependence of $\gamma(T)$ and $\eta(T)$ captures the transition between the signals coming from SC and the pseudogap phases and the curves of the two parameters cross at the critical temperature $T_c$=97 K for "nodal" measurements at low fluence (Figures 2b and d).

Figs. 2 **b** and **d** confirm that the overall transient response in the nodal region of the Brillouin Zone is not affected by the pump photon energy. On the contrary, in the antinodal region (Figures 2 **f** and **h**), the crossing temperature between the fit parameters strongly depends on the photon energy. In particular, for high pump photon energy the pseudogap contribution dominates, whereas for low photon energies the SC contribution prevails for some K above $T_c$. This reveals that the region in momentum space relevant to the onset of superconducting coherence above $T_c$ is the antinodal one, where the superconducting gap has its maximum value. Analysis for different excitation density (see supplementary) confirm this qualitative response.

In order to rationalize these results, we propose an effective d-wave BCS model interacting with a low-energy pump pulse (hν~2Δ) [see [20] for further details and *Supplemental Materials* for the symmetry analysis]. To compare the model with the experimental data, we computed the dynamics of the pair operator expectation value $\psi_k = \langle \hat{c}_\uparrow(\vec{k})\hat{c}_\downarrow(-\vec{k}) \rangle$ and tracked the abundance and the coherence of the pairs. In fact, in unconventional superconductors, the transition to the pseudogap phase is not determined by the vanishing of the pairing strength but of phase coherence [43-47]. In particular, we extracted two quantities: λ, the integral over the first Brillouin zone of the pair operator absolute value, which measures the abundance of pairs without considering their mutual and overall d-wave phase coherence, and ϕ, the absolute value of its integral, which measures the abundance of d-wave coherent pairs. Importantly, the contribution to the electronic response coming from excitation of different symmetry select specific regions (and signs for ϕ) in the Brillouin zone, we computed these quan0tities by integrating with (the absolute value in the λ case) $A_{1g}$, $B_{1g}$ or $B_{2g}$ kernels (see Fig 1 a, c and e) and simulate the results of the related symmetry-selective experiments.

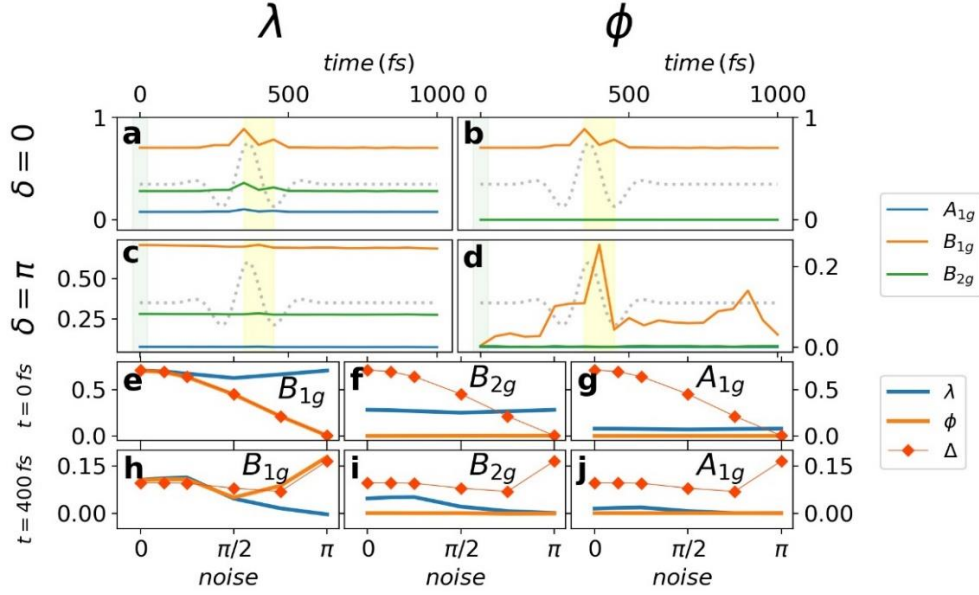

**Figure 3: Theoretical Effective model.** In **a** and **b**, the dynamics of λ and ϕ (see main text) are displayed. The blue line represents the $A_{1g}$ selection, the orange line the $B_{1g}$ one and the green curve the $B_{2g}$ one. The dashed line shows the time evolution of the pump pulse. In **c** and **d**, random noise with amplitude $\delta = \pi$ was added to the pair operator phase, in order to simulate an initial pseudogap state. In **e-j** the noise dependence of three quantities λ in blue, ϕ in orange and the superconducting gap with the red diamonds) is analyzed for two time delays: at the equilibrium (**e-g**) and immediately after the pump excitation (**h-j**).

The curves in Figure 3a show at time t=0 an initial number of pairs that depends on the symmetry selection as well as a time-dependent increase of their number that is correlated with the pump excitation (t=350 fs). The calculations reveal an enhancement of the number of pairs much stronger in the antinodal region in agreement with the experiment. A similar dynamical response is shown by the different contributions to the SC gap in Figure 3b, which unveil the prevalent effect of the preservation of the phase coherence over the increase of the number of pairs.

In order to describe the dynamical response above $T_C$, we propose to modify the equilibrium state of the BCS superconductor by introducing "artificially" phase incoherence, which could effectively mimic the presence of inhomogeneities in the system. This choice is justified by the evidence that in this temperature range SC fluctuations may persist, while macroscopic superconducting coherence is lost [46-53], likely because different spatial regions of the material can exhibit different pair phases [2,4,15]. In the calculations, we add a uniform random noise to the phase of the expectation value of the pair operator [20]. The dependence of the amplitude of the SC gap on the noise amplitude δ is shown in Fig. 3e (red diamonds), where the noise increase determines the complete collapse of the superconducting gap (for δ=π), despite the presence of pairs. Fig. 3 **e-g** show also that at the equilibrium the dominant contribution to the gap amplitude comes from the antinodal region (orange line in Fig. 3e) for any coherence condition.

By considering this initial arbitrary state with maximum incoherence (δ=π) and letting it evolve through our modified BCS Hamiltonian, we obtain the dynamical response for the different symmetries shown in Fig. 3 c-d. In this scenario, the pairs' abundance is much less affected by the pump excitation for any geometry (Fig. 3 c). On the contrary, a different behavior is observed for the SC gap: in the $B_{2g}$ and $A_{1g}$ configurations we observe a negligible modification of the gap induced by the pump, whereas the $B_{1g}$ contribution from the initial zero value (t=0 fs) increases sharply around t=300 fs (Fig. 3 c). These results suggest that a low photon energy excitation is able to dynamically restore phase coherence in the antinodal region of the system.

In conclusion, we studied the dynamical response of electronic excitations in optimally doped Y-B2212. By properly combining the polarization of the incoming and scattered probe beam, we were able to isolate the contributions to the pump wavelength dependence of the SC response observed in [20], coming from different regions of the Fermi Surface. In contrast to the nodal signal, the response at the antinodes is strongly dependent on the pump photon energy, while the nodal is not. More specifically, our results indicate that low photon energy excitations polarized along the Cu-Cu direction affect the of Cooper pairs density only slightly (and do not above $T_c$), but they dynamically induce pair coherence in the antinodal region (both above and below $T_c$). In the presence of an inhomogeneous system above $T_c$, this could give rise to a non-zero superconducting gap and superconducting-like behaviors, which might contribute to the observed light-driven coherent transport above Tc [54-61]. Finally, we stress that the possibility to enhance d-wave phase coherence at the antinodes through driven currents in the Cu-Cu direction may be of relevance for understanding both equilibrium and non-equilibrium properties of unconventional superconductors.

**Acknowledgments** This work was mainly supported by the European Commission through the ERCStG2015, INCEPT, Grant No. 677488. This research was undertaken thanks in part to funding from the Max Planck UBC Centre for Quantum Materials and the Canada First Research Excellence Fund, Quantum Materials and Future Technologies Program. The work at UBC was supported by the Killam, Alfred P. Sloan, and Natural Sciences and Engineering Research Council of Canada (NSERC) Steacie Memorial Fellowships (A.D.), the Alexander von Humboldt Fellowship (A.D.), the Canada Research Chairs Program (A.D.), NSERC, Canada Foundation for Innovation (CFI) and CIFAR Quantum Materials Program. The work at the University of Minnesota was funded by the Department of Energy through the University of Minnesota Center for Quantum Materials under DE-SC-0016371. A. A. acknowledges support by MIUR under Project PRIN 2017RKWTMY

# Supplemental Material

1. **EXPERIMENTAL DESIGN**

   **Laser System**

   In our experimental set-up the light source is provided by a Light Conversion Pharos Laser, producing 400 µJ/pulse with 1.2 eV photon energy at 50 KHz repetition. It pumps a Non-Collinear Parametric Amplifier (Orpheus-N by Light Conversion) and a Twin Optical Parametric Amplifier (Orpheus TWIN by Light Conversion), which generates the probe and pump pulse respectively.

   The optical probe is a 200 fs pulse, tunable in the visible (measurement reported at 1.63 eV). The MIR pump pulses are produced by Difference Frequency Generation (DFG) mixing the signal outputs of the twin OPA.

   **Sample**

   The sample is a large and high-quality optimally doped Y-substituted Bi2212 single crystals ($Bi_2Sr_2Y_{0.08}Ca_{0.92}Cu_2O_{8+\delta}$.) grown in an image furnace by the traveling-solvent floating-zone technique with a non-zero Yttrium content, as described in [1]. The sample critical temperature is Tc=97 K.

   **Derivation of the Raman matrix element**

In order to derive equation 1, let use consider the Raman tensor of associated to the $D_{4h}$ tetragonal point group. The Raman tensor can be decomposed in the basis of the irreducible representation of the symmetry group of the crystal as

$$R = \sum_n a_n R_{\Gamma_n} \quad (1)$$

where $\Gamma_n$ is a representation of the symmetry group [2].
For parity reasons, the only Raman active modes of the group are $A_{1g}$, $A_{2g}$, $B_{1g}$, $B_{2g}$ and $E_g$ [3,4] and, considering our experimental geometry, where the beam propagation direction is fixed and perpendicular to the Cu-O plane, the resulting Raman tensor is [4]:

$$R_{B12212} = \frac{1}{2}\begin{bmatrix} R_{A1g} + R_{B1g} & R_{B2g} \\ R_{B2g} & R_{A1g} - R_{B1g} \end{bmatrix} \quad (2)$$

Let us introduce the angles $\alpha$ and $\theta$, which represent the polarization directions with respect to the crystallographic Cu-O axis of the input and output beam respectively; the matrix element of the Raman tensor $R_{\mu\nu}=R(\alpha,\theta)$ is calculated on two states, that is, the incoming beam with polarization $\hat{\varepsilon}_I \propto \begin{pmatrix} \cos\alpha \\ \sin\alpha \end{pmatrix}$ and the reflected beam $\hat{\varepsilon}_F \propto \begin{pmatrix} \cos\theta \\ \sin\theta \end{pmatrix}$. Applying the two states to the matrix 4, one easily obtains the relation 1.

**Detection**

Looking at Equation 1, it is clear that the choice of the orthogonal angles $\alpha$ and $\theta$ is not the only way to obtain $B_{1g}$ and $B_{2g}$ dynamics. As a matter of fact, the subtraction of two signals with orthogonal $\theta$ angles can produce the same result. For example, for $\alpha=0°$ and $\theta_{1,2}=\pm 45°$ one gets

$$R(\alpha,\theta_{1,2}) = \frac{1}{\sqrt{2}}\left[R_{A1g} + R_{B1g} \pm R_{B2g}\right] \quad (3)$$

and so $R(\alpha,\theta_1) - R(\alpha,\theta_2) \propto R_{B2g}$ (see Figure S1 and S2).

Performing the same calculation for $\alpha=45°$, $\theta_1=0°$ and $\theta_2=90°$, one gets $R(\alpha,\theta_1) - R(\alpha,\theta_2) \propto R_{B1g}$.

This demonstrates that birefringent measurements, i.e. the detection of the difference between two orthogonal components of the probe pulses, with suitable angles can represent an alternative technique to acquire $B_{1g}$ and $B_{2g}$ signals, as illustrated in Figure S1. In order to check the validity of the alternative method, in Figure S2 we compare the $B_{2g}$ dynamics measured with the two techniques: the orange line is obtained with the standard configuration ($\alpha=0°$ and $\theta=90°$) whereas the black one refers to the birefringence measurement. The two signals are proportional both in the superconducting phase (Figure S2 a and c) and in pseudogap (Figure S2 b and d) and in the two fluence regimes (linear: Figure S2 a and b, non-linear: c - d).

The measurement of the birefringence is particularly convenient because the cross polarization measurements needed to select $B_{1g}$ and $B_{2g}$ signals can be disturbed by $A_{1g}$ contributions, often present because of the high intensity of the total symmetric signal with respect to the others. On the contrary, in a birefringence measurement the $A_{1g}$ component is split in two contributions, which are equal if the polarization angles $\theta_{1,2}$ are symmetric with respect to the probe polarization $\alpha$. The final subtraction cancels completely the total symmetric contribution.

So, we performed birefringence time resolved measurements with suitable input and output polarization directions ($(\alpha,\theta_{1,2}) = (0°, \pm 45°)$ and $(\alpha,\theta_{1,2}) = (45°, \pm 0 - 90°)$)

The orthogonal polarization components of the probe are split by a Wollaston prism after the interaction with the sample. The final signal is the subtraction of the two projections of the probe beam, performed by a differential photodetector. In the configuration needed to select $B_{1g}$ and $B_{2g}$ symmetries the prism is always oriented such that the measured signal at the equilibrium (t<0) has the same projection on the two axes.

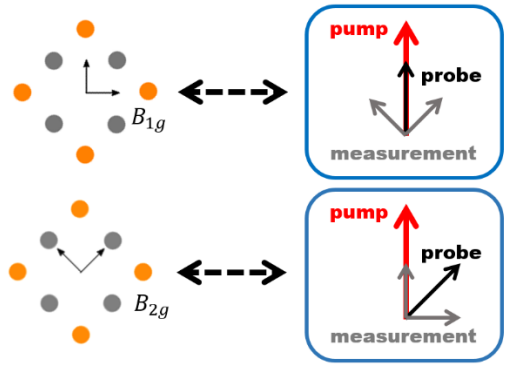

**Figure S1: Measurement correspondence**. Correspondence between ``standard'' pump-probe polarization configurations used to measure B1g and B2g signals (on the left) and birefringence measurements (on the right). The colored dots on the left represent the copper (orange) and oxygen (grey) atoms, whereas the arrows show the polarization of the incoming probe beam and the polarization selection after the sample. On the right panel the polarizations of the birefringence measurement are shown: the black line represents the incoming probe and the grey ones the two final polarization selections after the sample.

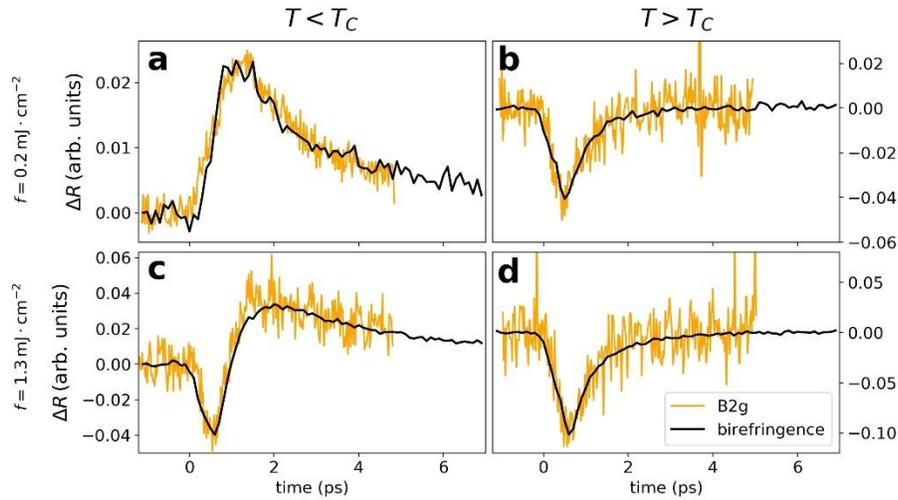

**Figure S2: Measurement comparison.** Comparison between a pump probe measurement performed in $B_{2g}$ configuration (orange line) and a birefringence measurement (black line, obtained by subtracting the polarization components along the Cu-Cu axes), both in superconducting phase (**a** and **c**) and in pseudogap (**b** and **d**). The measurements have been performed in two different pump fluence regimes (0.2 mJ/cm$^2$ in **a** and **b** and 1.3 mJ/cm$^2$ in **c** and **d**). The signals have been rescaled in order to verify the proportionality between the two measurements.

## 2. FITTING PROCEDURE

In order to check the validity of the fitting procedure described in the main text (Equation 2), here we analyze the results obtained for higher fluence excitations and finally compare the measured data with the results of the fit.

### High fluence measurements

Figure S3 shows the temperature maps of the pump probe measurements in which the sample is excited with higher fluence pulses ($f = 0.4\ mJ \cdot cm^{-2}$), in order to check the temperature dependence of the parameters A and B of the fitting procedure. In particular, the results reported in Figure S3 b, d, f and h suggest that the crossing between the two parameters, which should occur at the transition temperature in our interpretation, is shifted to lower temperatures with respect to the low fluence measurements reported in the main text. We interpret this behavior as the result of the strong non-linear excitation produced by the pump pulse, which is able to dynamically perturbed (or even destroy) the superconducting phase, resulting in an effective decrease of the measured Tc. Interestingly, the effect is much more visible when we select the response of the antinodal region ($B_{1g}$ configuration, Figure S3 f and h), where the SC gap has its maximum value at the equilibrium, with respect to the nodal one ($B_{2g}$ symmetry, Figure S3 b and d).

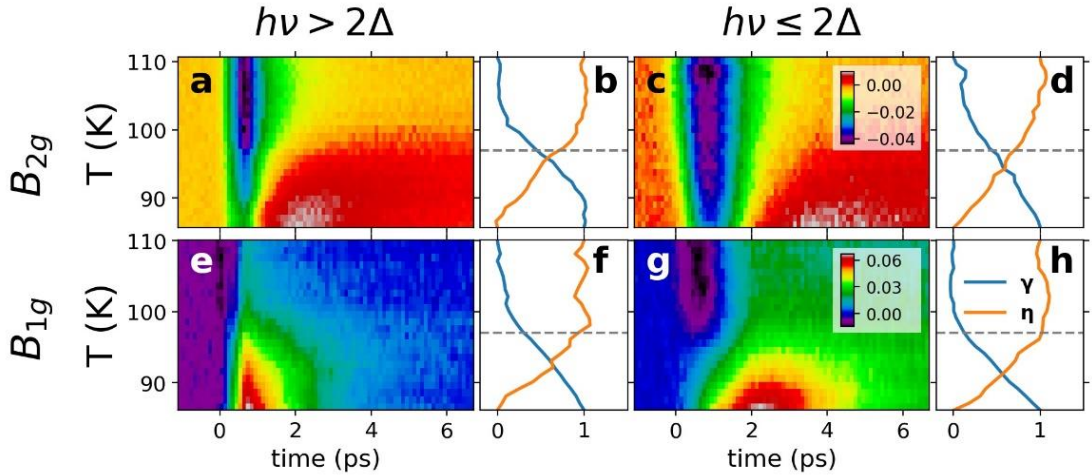

**Figure S3: High fluence measurements.** Temperature-dependent pump-probe measurements, in $B_{2g}$ and $B_{1g}$ configurations, with high fluence excitations (0.4 mJ cm$^{-2}$). First row: $B_{2g}$ measurements with different excitation photon energies (**a** 170 meV and **c** 70 meV). The graphs **b** and **d** display the values of $\gamma(T)$ and $\eta(T)$ (Eq. 6) obtained from the data in **a** and **c**, respectively. The dashed grey lines represent the critical temperature T$_c$. The bottom row (figs. e-h) shows the same analysis for measurements in the $B_{1g}$ configuration.

### Results of the fitting procedure

In Figure S4 and S5 we report the complete outcome of the fitting process for nodal ($B_{2g}$) and antinodal ($B_{1g}$) response selection respectively, in order to compare the results of the analysis with the linear fit and to check the validity of the procedure.

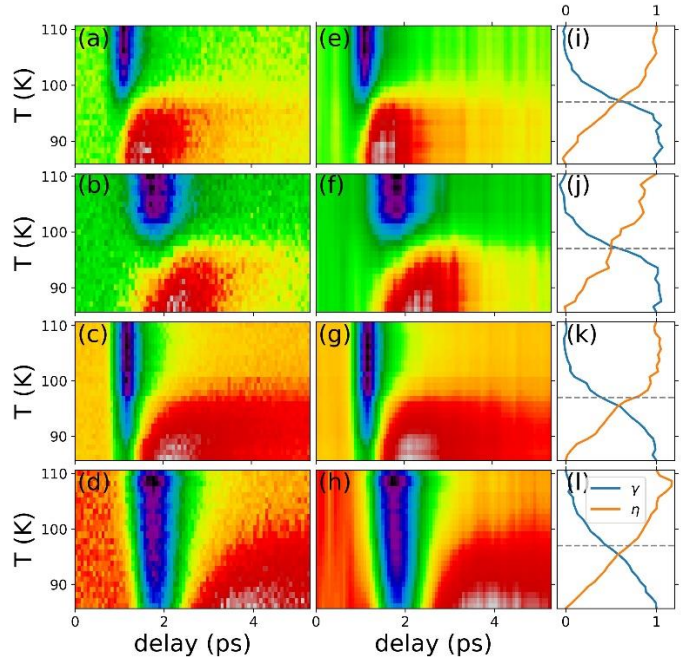

**Figure S4: Comparison between data and fit – B$_{2g}$ symmetry.** In the left column the measured data are plotted (in the order nodal dynamics for low (a-b) and high (c-d) excitation fluence and high (a, c) and low (b,d) photon energy). In the central column the corresponding fitted maps are shown. The right column displays the values of the fitting parameters $\gamma$ and $\eta$ as a function of the temperature. The dashed grey lines represent the critical temperature T$_c$.

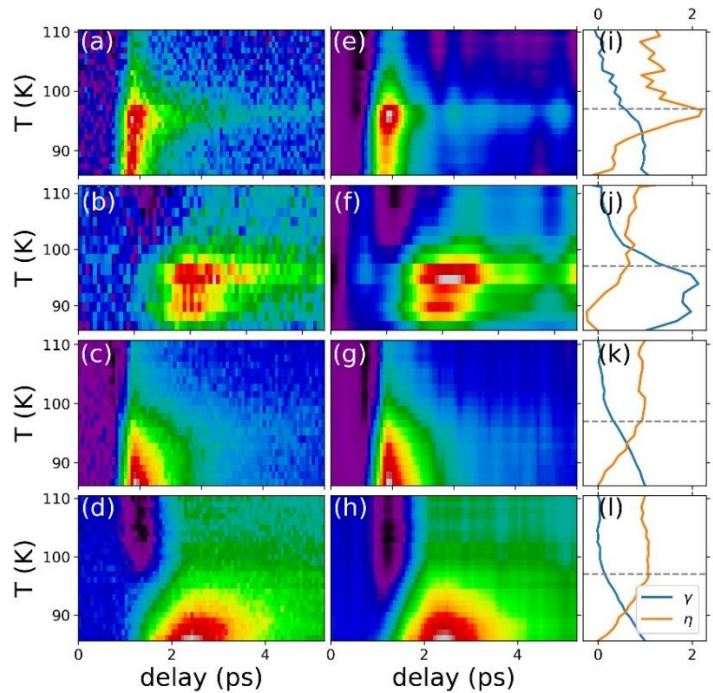

**Figure S5: Comparison between data and fit – B$_{1g}$ symmetry.** In the left column the measured data are plotted (in the order nodal dynamics for low (a-b) and high (c-d) excitation fluence and high (a, c) and low (b,d) photon energy). In the central column the corresponding fitted maps are shown. The right column displays the values of the fitting parameters $\gamma$ and $\eta$ as a function of the temperature. The dashed grey lines represent the critical temperature T$_c$.

## 3. BCS EFFECTIVE HAMILTONIAN

In order to interpret the dynamics of the signal, we consider a modified BCS Hamiltonian, which takes into account also the excitation produced by the pump pulse and the anisotropic SC gap, characteristic of high temperature superconductors.

This Hamiltonian can be written as

$$H = \sum_k \varepsilon\left(\vec{k} - \frac{e}{\hbar}A(t)\vec{\chi}\right)\hat{n}(\vec{k}) + \sum_k \left(\Delta^*(\vec{k})\vec{\psi}_k + \Delta(\vec{k})\vec{\psi}_k^+\right) \quad (4)$$

where $\varepsilon(\vec{k})$ is the two dimensional tight binding electronic dispersion, $A(t)$ is the vector potential of the pump pulse and $\epsilon$ is its polarization. Finally $\Delta(\vec{k})$ is the gap function, whose dependence on k remarks its d-wave anisotropy and $\hat{\psi}(\vec{k})$ is the pair operator.

The Hamiltonian is used to calculate the time dependence of the quantity named in the main text, that allows to check the dynamics of the number of Cooper pair and of their phase coherence after the pump excitation. For further details, see [5].